\documentclass[5p,times,twocolumn]{elsarticle}
\usepackage{amsmath,amssymb}
\biboptions{comma,sort&compress}
\usepackage[utf8]{inputenc}
\usepackage{graphicx}% Include figure files
\usepackage{bm}% bold math
\usepackage{hyperref} % add hypertext capabilities
\hypersetup{colorlinks=true, linkcolor=blue, anchorcolor=red, citecolor=blue, filecolor=red, urlcolor=red, pdfauthor=author}
\usepackage[dvipsnames]{xcolor}

\usepackage{dblfloatfix}

\usepackage{etoolbox}
\makeatletter
\patchcmd{\abstract}{\par \medskip}{\par \smallskip}{}{}
\patchcmd{\abstract}{\noindent \unskip \textbf {\@elsarticleabstitle }}{\vspace{-0.4\baselineskip}\noindent\unskip\textbf{\@elsarticleabstitle}}{}{}
\makeatother

\journal{Physics Letters B}

\begin{document}

\begin{frontmatter}

\title{Bayesian model-data comparison incorporating theoretical uncertainties}

\author[osu]{{Sunil Jaiswal}\corref{cor}}
\ead{jaiswal.61@osu.edu}
\author[wsu]{Chun Shen}
\ead{chunshen@wayne.edu}
\author[osu]{Richard J. Furnstahl}
\ead{furnstahl.1@osu.edu}
\author[osu]{Ulrich Heinz}
\ead{heinz.9@osu.edu}
\author[isu]{Matthew T. Pratola}
\ead{mpratola@iu.edu}

\cortext[cor]{Corresponding author}

\address[osu]{Department of Physics, The Ohio State University, Columbus, Ohio 43210, USA}
\address[wsu]{Department of Physics and Astronomy, Wayne State University, Detroit, Michigan 48201, USA}
\address[isu]{Department of Statistics, Indiana University, Indiana 47405, USA}

\date{\today}

%%%%%%%%%%%%%%%%%%%%%%%%%%%%%%%%%%%%%%%%%%%%%%%%%%%%%%%%%%%%%%%%

\begin{abstract}
Accurate comparisons between theoretical models and experimental data are critical for scientific progress. However, inferred physical model parameters can vary significantly with the chosen physics model, highlighting the importance of properly accounting for theoretical uncertainties. In this Letter, we present a Bayesian framework that explicitly quantifies these uncertainties by statistically modeling theory errors, guided by qualitative knowledge of a theory's varying reliability across the input domain. We demonstrate the effectiveness of this approach using two systems: a simple ball drop experiment and multi-stage heavy-ion simulations. In both cases incorporating model discrepancy leads to improved parameter estimates, with systematic improvements observed as additional experimental observables are integrated.
\end{abstract}
%%%%%%%%%%%%%%%%%%%%%%%%%%%%%%%%%%%%%%%%%%%%%%%%%%%%%%%%%%%%%%%%

\end{frontmatter}

%-------------------------------------------
\section{Introduction} \label{sec:intro}
\vspace{-.1cm}
%-------------------------------------------

Comparisons between theoretical models and experimental data are at the heart of scientific inquiry. Theoretical models guide our understanding of complex systems by translating hypotheses into quantitative predictions that can be tested experimentally. Traditionally, a close fit between a model's predictions and measured data is interpreted as a sign of success, often implying that the model parameters capture the underlying physical processes. However, this paradigm assumes that the model fully represents the complexity of actual systems -- an assumption that is rarely justified in practice. All models have inherent limitations beyond their domains of validity, and using them beyond these regimes without accounting for theoretical uncertainties can lead to biased parameter estimates, reducing these parameters to mere ``fitting variables'' rather than meaningful physical quantities~\cite{Kennedy2002}. Moreover, discrepancies between certain measurements and otherwise successful models sometimes lead researchers to assign lower weights to these data, thereby diminishing their utility and limiting the potential insights they can provide.

Quantifying theoretical uncertainties remains a significant challenge, particularly in systems with intricate interactions or multistage dynamics. A striking example are relativistic heavy-ion collisions where nuclei collide at near-light speeds and produce quark-gluon plasma (QGP) \cite{Gyulassy:2004zy, Harris:2023tti}. The evolution of this deconfined matter involves several stages --- initial energy deposition, hydrodynamic expansion, hadronization, and freeze-out --- each introducing additional layers of assumptions and uncertainties into the model~\cite{Gale:2013da, Shen:2020mgh, Heinz:2024jwu}. Without a systematic framework to disentangle true physical effects from model inadequacies, parameter extraction risks leading to overfitting rather than genuine discovery.

A model discrepancy framework employing Gaussian processes (GPs) was introduced in \cite{Kennedy2002}, and variations have been explored in many studies \cite{Higdon2004, Bayarri01052007, Arendt:etal2012:1, Arendt:etal2012:2, Brynjarsdottir_2014, GARDNER2021107381}. In these approaches, the discrepancy between experimental data and theoretical model predictions, stemming from missing physics or approximations, is modeled using a GP. However, one persistent challenge is the need to constrain the GP's covariance kernel. For example, in \cite{Brynjarsdottir_2014}, the authors emphasized the importance of incorporating knowledge of the theory's validity at specific points in the input space (i.e., the domain in which observables are measured) so that both the GP and its derivative could be accurately constrained. In practice, however, specifying such accurate knowledge about the theory is often difficult. In this Letter, we construct the GP covariance kernel based on only qualitative prior knowledge of the theory's domain of validity across the input space. This type of knowledge -- for example, recognizing that ``\textit{the theory is more reliable in this regime than in that one}'' -- is typically easier to provide and often available. By leveraging this information, the framework prioritizes the accurate extraction of model parameters rather than simply optimizing the fit to the observables. We perform Bayesian parameter inference to simultaneously estimate both the model parameters and the GP hyperparameters, thereby quantifying uncertainties from both the experimental data and the theoretical model.

We test the framework on two systems, a motivating ball-drop experiment and various multi-stage simulations of heavy-ion collisions. By comparing our results with those obtained using Bayesian inference that does not account for theoretical uncertainties, we show that incorporating a constrained model discrepancy into the statistical analysis yields parameter estimates that are more robust and closer to the true values. Furthermore, the accuracy of the inferred model parameters is shown to improve systematically when increasing the number of experimental observables.

%%%%%%%%%%%%%%%%%%%%%%%%%%%%%%%%%%%%%%%%%%%%%%%%%%%%%%%%%%%%%%%%

%-------------------------------------------
\section{Framework} \label{sec:framework}
%-------------------------------------------

We begin by setting up the model discrepancy framework following \cite{Kennedy2002}. Let the $i$-th measurement for an observable $y$ be denoted as
 \begin{equation}\label{eq:stat_true}
    y(x_i) = \zeta(x_i) + \epsilon_i \,,
    \qquad \epsilon_i \sim \mathcal{N}(0,\sigma_i^2) \,.
 \end{equation}
Here, the index $i$ corresponds to the point in input space $x$ where the $i$-th observation is made (e.g., a specific time in the ball drop experiment or a particular transverse momentum $p_T$ in the heavy-ion simulation discussed later). Independent observation errors $\epsilon_i$ are assumed to follow Gaussian distributions with zero mean and standard deviation $\sigma_i$. Here, $\zeta(x_i)$ represents the true value of the observable at $x_i$.\footnote{%
    We assume that $x$ is made unitless by scaling it with a suitable reference scale $x_0$.
    }%

We denote the prediction from a theoretical model for the observable $y$ as $\eta(x, \bm\theta)$, where $\bm\theta$ is the vector of true but unknown model parameters. The model for $\eta(x,\bm\theta)$ is then defined as:
 \begin{equation}\label{eq:md}
    \zeta(x) = \eta(x, \bm{\theta}) + \delta(x) \,;
 \end{equation}
here $\delta(x)$ quantifies the discrepancy between the experimental data and the theoretical model prediction with the true model parameters at input setting $x$. Combining the above equations~\eqref{eq:stat_true} and \eqref{eq:md}, we express the observation $y(x_i)$ as
 \begin{equation}\label{eq:stat_MD}
    y(x_i) = \eta(x_i, \bm{\theta}) + \delta(x_i) + \epsilon_i \,.
 \end{equation}
Thus, each observation is modeled as the sum of the model output (evaluated at the true $\bm\theta$), the model discrepancy $\delta$ at $x_i$, and the observational error $\epsilon_i$. 

If the functional form of $\delta(x)$ were known exactly, fitting the corrected model $\eta(x, \bm{\theta}) + \delta(x)$ to data would, in principle, yield correct estimates of $\bm\theta$. However, $\delta(x)$ is usually not known precisely, and at best only qualitative knowledge about this error is known. We therefore treat $\delta(x)$ statistically and model it with a Gaussian process that incorporates this knowledge of the theory error.

Following Kennedy and O'Hagan~\cite{Kennedy2002}, we represent $\delta(x)$ as a zero-mean Gaussian process (GP) \cite{Rasmussen2004}:
 \begin{equation}
    \delta(\cdot \mid\bm\phi) \sim {\rm GP} \left(\bm 0, K(\cdot,\cdot \mid\bm\phi) \right) \,,
 \end{equation}
where $K(\cdot,\cdot \mid\bm\phi)$ is the covariance kernel, which depends on a set of hyperparameters $\bm\phi$. A GP represents a distribution over functions, with the covariance kernel $K(\cdot,\cdot \mid\bm\phi)$ serving as the central object that defines the properties of the functions in the distribution. In this framework, we incorporate prior knowledge about the error of the theory through the choice of the covariance kernel. This prior information mitigates the identifiability problem between the theoretical model and the model discrepancy, as discussed in detail in Ref.~\cite{Brynjarsdottir_2014}.
Adopting a Bayesian approach, we assign prior probability distributions to both the model parameters $\bm\theta$ and the GP hyperparameters $\bm\phi$, and update these to obtain their posterior distributions conditioned on the observations. In all examples discussed in this work, uniform priors (in appropriate ranges) are used for both the model parameters and the GP hyperparameters.

%%%%%%%%%%%%%%%%%%%%%%%%%%%%%%%%%%%%%%%%%%%%%%%%%%%%%%%%%%%%%%%%

%-------------------------------------------
\subsection{Choice of covariance kernel} 
\label{sec:kernel}
%-------------------------------------------

In the examples explored in this work, we consider two different covariance kernels for the discrepancy GP, each encoding distinct prior knowledge about our confidence in the model's validity across different values of $x$.%
    \footnote{The choice of covariance kernel for the discrepancy GP should reflect prior knowledge about theory error. In the examples considered here, we assume that the only information known about the theory error is that it increases as a function of the input space, and thus the kernels are constructed accordingly. If more detailed information about the theory error is available, a more precise kernel can be constructed} to incorporate this knowledge explicitly.
%%%%%%%%
\begin{enumerate}
    \item Kernel \texttt{I}: 
    \begin{equation}\label{kernel1}
        K(x_i, x_j \mid\bm\phi) \equiv 
        s^2 + \bar{c}^2 (x_i x_j)^r \exp\left( -\frac{\lVert x_i-x_j\rVert^2}{2\ell^2} \right).
    \end{equation}
    This kernel encodes the prior belief that \textit{the model is more reliable at small $x$ than at large $x$}. The term $s^2$ represents a baseline variance that is present across all $x$. The power term $(x_i x_j)^r$ increases with $x$ (for $r\geq 0$), amplifying the magnitude of the variance at larger $x$. This reflects increasing uncertainty in the model at large $x$, incorporating our prior belief about the theory error. The exponential decay term $\exp(-\frac{\lVert x_i - x_j\rVert^2}{2\ell^2})$ ensures that discrepancies remain correlated over short distances in $x$, but correlations decay as $x_i$ and $x_j$ become further apart. Finally, the parameter $\bar{c}^2$ scales the overall magnitude of the second term, determining the strength of the discrepancy relative to the baseline term $s^2$. This kernel has four hyperparameters: $\bm \phi = (s, \bar{c}, r, \ell)$.

    \item Kernel \texttt{II}:
    \begin{equation}\label{kernel2}
        K(x_i, x_j \mid\bm\phi) \equiv \bar{c}^2 (x_i x_j)^r \exp\left( -\frac{\lVert x_i-x_j\rVert^2}{2\ell^2} \right).
    \end{equation}
    This kernel reflects the assumption that \textit{the model is correct at $x=0$, but our confidence in its validity decreases as $x$ increases}. Unlike Kernel \texttt{I}, it does not include a baseline variance term $s^2$, implying complete confidence in the model at $x=0$. This kernel has three hyperparameters: $\bm \phi = (\bar{c}, r, \ell)$.

\end{enumerate}
%%%%%%%%

%--------------------------------
\begin{figure*}[!t]
    \centering
    \includegraphics[width=\linewidth]{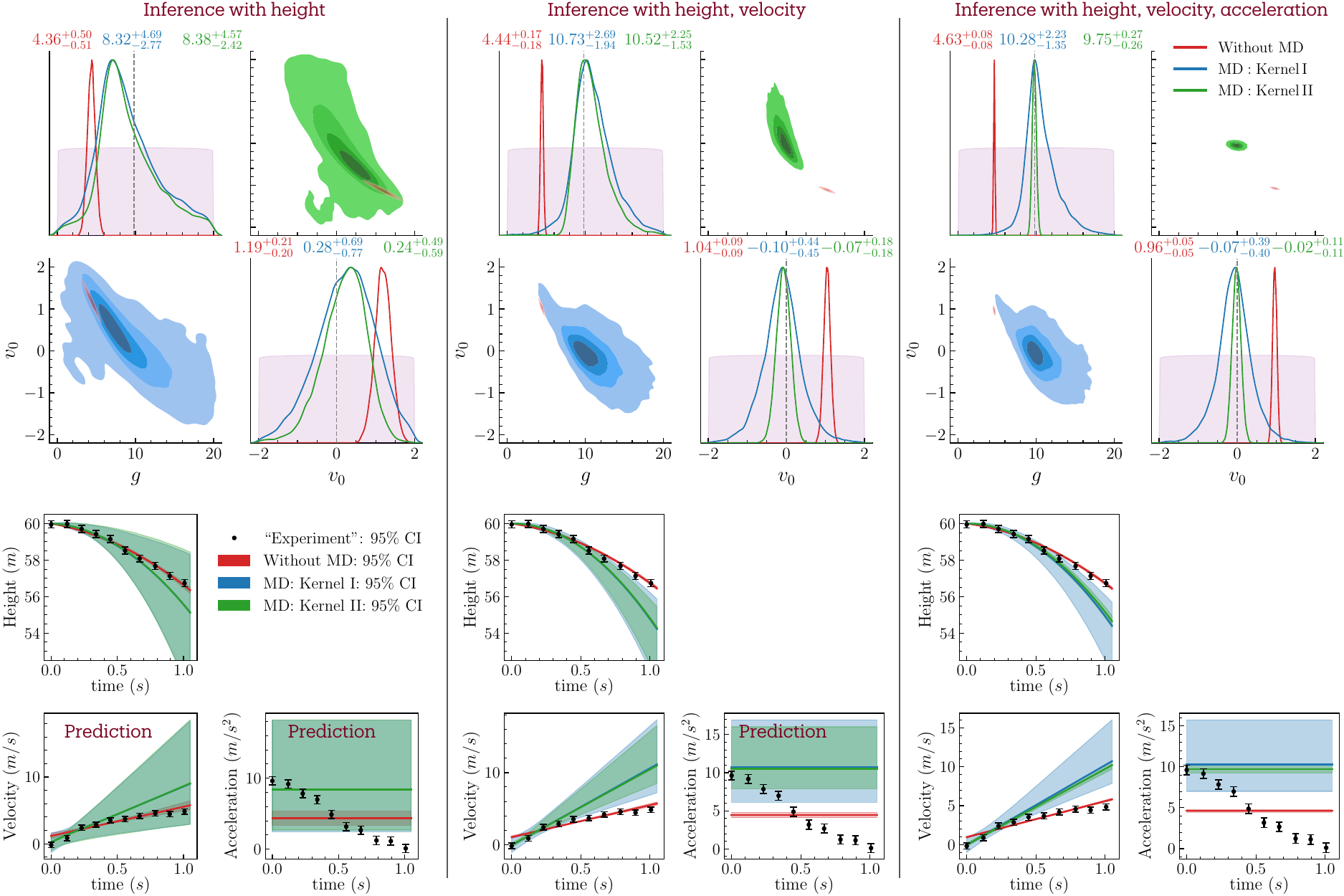}
    \vspace{-6mm}
    \caption{Results of Bayesian inference for the ball drop experiment. 
    \textbf{Top row:} Corner plots displaying the inferred posteriors for $g$ and $v_0$ (with median $\pm 68\%$ CI displayed on the diagonals) as additional observables (specified above the panels) are sequentially incorporated (from left to right) into the Bayesian inference. \textbf{Bottom row:} Model predictions based on the inferred parameters corresponding to the cases in the top row.}
    \vspace{-2mm}
    \label{fig:balldrop}
\end{figure*}
%--------------------------------

Note that we use separate discrepancy GPs, each with its own set of hyperparameters, for different observables. For the examples discussed in this work, both Kernel \texttt{I} and Kernel \texttt{II} are appropriate choices.

At this point, we explain how this framework enables accurate inference of physical parameters. Incorporating the discrepancy term in Eq.~\eqref{eq:stat_MD} effectively assigns weights to the model parameters. In this approach, the discrepancy $y(x_i) -\eta(x_i, \bm\theta^*)$, with $\bm\theta^*$ representing the true parameter values, is modeled using prior theoretical knowledge such that greater weight is given to $\bm\theta^*$ at small values of $x$. This occurs because the GP generates a distribution over functions that complies with the constraints specified by the covariance kernel. For example, the kernels considered above are designed to generate only functions whose magnitude can increase with $x$, thereby emphasizing the true model parameters $\bm\theta^*$. Conversely, parameters $\bm\theta$ for which the discrepancy between the data and the model predictions does not increase with $x$ are assigned less weight. The kernel hyperparameters (particularly the power $r$), which are estimated together with the model parameters, determine the rate at which the discrepancy increases with $x$.

%%%%%%%%%%%%%%%%%%%%%%%%%%%%%%%%%%%%%%%%%%%%%%%%%%%%%%%%%%%%%%%%

%-------------------------------------------
\section{Ball drop experiment}  \label{sec:ball_drop}
%-------------------------------------------

We test the model discrepancy framework first on a simple system in which a ball is dropped from a tower of height $h_0=60\,$m at time $t=0\,$s~\cite{Higdon:etal:2010}. Measurements of the ball's height above the ground, velocity, and acceleration are recorded at discrete time points until $t=1$\,s.\footnote{%
	In this example $x=t$/(1\,s) serves as the unitless input parameter, $0\leq x \leq 1$.}
The true theory, used to generate mock experimental data, incorporates a gravitational force $\mathbf{f}_G=-Mg\,\hat{\bf z}$ and a drag force given by $\mathbf{f}_D = -0.4 v^2\, \hat{\bf{v}}$. Here, $M=1\,$kg is the mass of the ball, $g=9.8\,{\rm m/s}^2$ the acceleration due to gravity, and we set the initial velocity $v_0$ to $v_0=0\,$m/s.

We assume that the observational errors for height, velocity, and acceleration are independent and identically distributed Gaussian random variables with standard deviations of $\sigma = 0.1, 0.2$ and $0.3$, respectively. Mock experimental data are generated by sampling from Gaussian distributions with these standard deviations and with means given by the predictions of the true theory, which are determined by the equations of motion:
\begin{equation}\label{eq:BD_evol}
    a = \frac{d v}{dt} = g - 0.4 v |v| \,, \qquad 
\frac{dh}{dt} = -v \,.
\end{equation}

We compare these mock data with a theoretical model that neglects the drag force and is governed by the simpler equations of motion:
\begin{equation}
    v = v_0 + g t \,, \qquad h = h_0 - v_0 t - \frac{1}{2} g t^2 \,.
\end{equation}
Here $h_0=60\,$m is fixed to the value used in generating the mock experimental data. The parameters $g$ and $v_0$ are model parameters, which we infer using Bayesian parameter inference. We incorporate our prior knowledge about the theory's domain using the two distinct GP kernels discussed earlier. Since our model neglects drag forces, we are more confident in its predictions at earlier times (small $x$), when the velocity is lower, than at later times (larger $x$).

Figure~\ref{fig:balldrop} shows both the inferred posteriors for the model parameters and the corresponding model predictions. We perform Bayesian parameter inference under three scenarios: (i) without the model discrepancy (MD) term (red), (ii) with MD using GP-kernel \texttt{I} (blue), and (iii) with MD using GP-kernel \texttt{II} (green). Parameter inference is carried out by sequentially incorporating additional observables to examine their influence on the inferred posteriors. In the top row of Fig.~\ref{fig:balldrop}, the left panel shows the corner plot when only height measurements are considered, the middle panel adds measurements of the velocity, and the right panel includes all observables. The vertical dashed gray lines in the diagonal panels mark the true parameter values used to generate the mock data, and the shaded purple regions denote the priors.

%--------------------------------
\begin{figure}[t!]
    \centering
    \includegraphics[width=\linewidth]{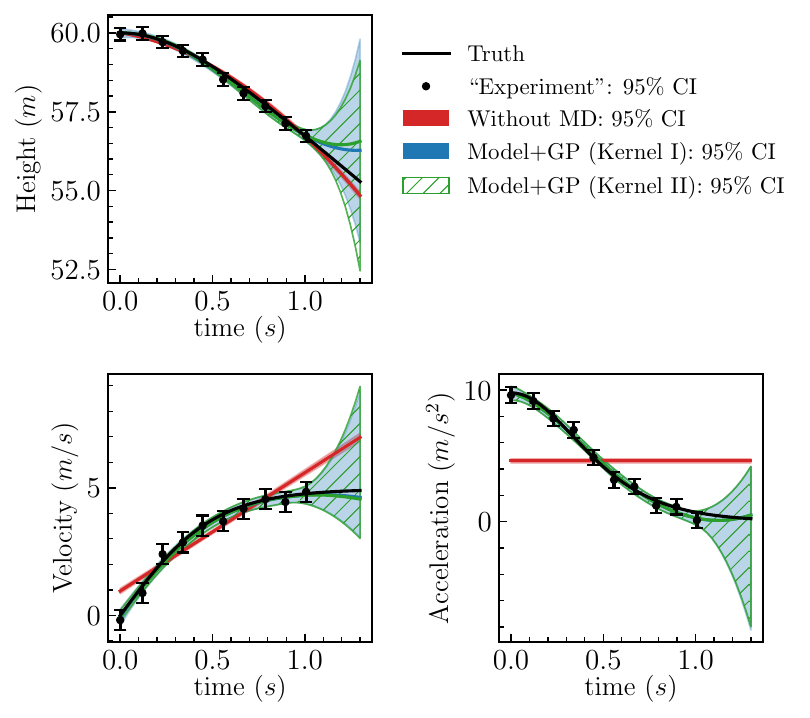}
    \vspace{-6mm}
    \caption{Predictions based on inferred parameters when all observables are considered in calibration. The black curve shows the prediction from the true theory. The blue band shows the predictions of model + discrepancy GP with Kernel I, and green shows the same with kernel II. The red band shows the model predictions without accounting for model discrepancy (same as in the bottom-right panel of Fig.~\ref{fig:balldrop})}
    \vspace{-2mm}
    \label{fig:BD_hva_model_wMD}
\end{figure}
%--------------------------------

The posteriors incorporating MD (blue and green) consistently cover the true values and narrows as more observables are included. In particular, the posteriors with MD using Kernel \texttt{II} (green), which encodes strong prior information about the theory's domain of validity, are consistently tighter, reflecting greater constraining power, yet remain consistent with those obtained using the more conservative Kernel \texttt{I} (blue). Remarkably, when acceleration data are included (right panel), the posterior with Kernel \texttt{II} nearly recovers the true parameter values, even though the model predicts a constant acceleration that does not capture the observed decreasing trend in the data. In contrast, the posteriors from the inference without the MD term (red) deviate significantly from the truth and become increasingly narrow as more observables are added, resulting in model parameter estimates that are both incorrect and overconfident.

In the bottom row of Fig.~\ref{fig:balldrop}, the mock experimental data are compared with the model predictions generated using the inferred parameter posterior from the top row. Notably, predictions with MD (blue, green) match the data at early times and deviate at later times, a behavior that directly results from incorporating prior knowledge about the theory's domain of applicability into the covariance kernel. In contrast, predictions without MD (red) prioritize achieving the best fit to the data, leading to inaccurate and overconfident parameter estimates.

We show the model + discrepancy GP predictions in Fig.~\ref{fig:BD_hva_model_wMD}, computed using the inferred model and GP parameters with all observables included in the calibration. The blue band shows the combined predictions with Kernel I, and the green band shows those with Kernel II. In both cases, the combined predictions agree with the observations, as expected from Eq.~\eqref{eq:stat_MD}, provided the discrepancy GP is sufficiently flexible to model the theoretical error. The predictions for velocity and acceleration remain consistent with the correct model (black curve) in extrapolation regions. For height, the predictions deviate from the correct model immediately outside the range of the observations. The uncertainties of the combined predictions for all observables increase rapidly in extrapolation regions because the discrepancy GP kernels include a variance envelope proportional to $(x_i x_j)^r$. A more theory-error informed kernel may extrapolate better. Extrapolation issues in the model discrepancy framework are discussed in detail in Ref.~\cite{Brynjarsdottir_2014}.

%%%%%%%%%%%%%%%%%%%%%%%%%%%%%%%%%%%%%%%%%%%%%%%%%%%%%%%%%%%%%%%%

%-------------------------------------------
\section{Heavy-ion simulation}  \label{sec:heavy_ion}
%-------------------------------------------

%--------------------------------
\begin{figure*}[tbh!]
    \centering
    \includegraphics[width=\linewidth]{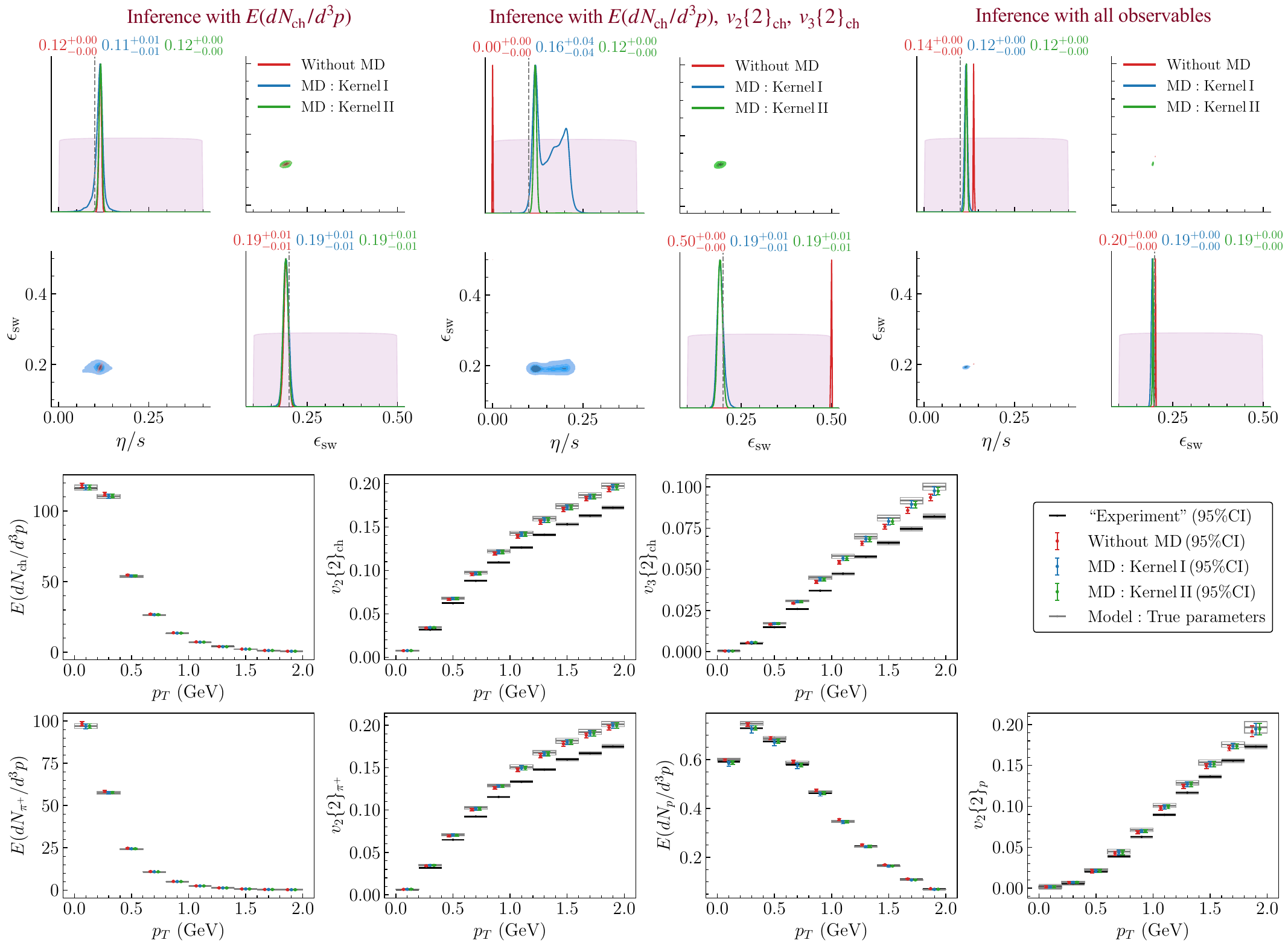}
    \vspace{-6mm}
    \caption{Results for the two-parameter hydrodynamic simulation using mock data generated with fixed $\eta/s=0.1$ and $\epsilon_{\rm sw}=0.2\,$GeV/fm$^3$. 
    \textbf{Top row:} Corner plots showing the posteriors for the inferred $\eta/s$ and $\epsilon_{\rm sw}$ (with median $\pm 68\%$ CI displayed on the diagonals) as more observables are included in the Bayesian inference (from left to right).
    \textbf{Bottom rows:} Model predictions based on inference using all observables.}
    \vspace{-2mm}
    \label{fig:2flat}
\end{figure*}
%--------------------------------

In the following examples we simulate collisions of Au on Au nuclei at $\sqrt{s_\textrm{NN}} =200\,$GeV at $20-30\%$ collision centrality, using the \texttt{iEBE-MUSIC} simulation framework \cite{Shen:2014vra,Shen:2022oyg}. Initial conditions are generated using the \texttt{MC-Glauber} model and coupled with relativistic viscous hydrodynamics \texttt{MUSIC}~\cite{Paquet:2015lta} assuming longitudinal boost-invariance, followed by hadronic transport using the \texttt{UrQMD} model \cite{Bass:1998ca, Bleicher:1999xi}. When the energy density in a cell drops below the value $\epsilon_{\rm sw}$, the fluid is ``particlized'', using the \texttt{iS3D} hadron sampler \cite{Shen:2014vra} which converts the fluid's energy and momentum into various hadron species $i$ with momenta ${\bf p}{\,=\,}({\bf p_T},p_L)$ using the Cooper-Frye prescription \cite{Cooper:1974mv}: 
\begin{align}
    E \frac{dN_i}{d^3 p} = \frac{g_i}{(2\pi)^3} \int_\Sigma d^3 \sigma_\mu p^\mu (f_\mathrm{eq} + \delta f)_i.
\end{align}
Here $\Sigma$ is the particlization hypersurface \cite{Huovinen:2012is}, characterized by a constant energy density $\epsilon_{\rm sw}$, $g_i$ is the spin degeneracy factor for particle species $i$, $f_\mathrm{eq}$ stands for the thermal equilibrium distribution and $\delta f$ is the out-of-equilibrium corrections. 

In the following study these simulations are considered to represent the ``truth''. We use them to generate mock experimental data by making measurements of the charged hadron multiplicity 
$E(dN_{\text{ch}}/d^3p)$, their elliptic and triangular flows $v_2\{2\}_{\rm ch}$ and $v_3\{2\}_{\rm ch}$, as well as the multiplicities $E(dN/d^3p)$ and elliptic flows $v_2\{2\}$ for identified pions and protons, in ten $p_T$ bins spanning the range $0{-}2\,$GeV.\footnote{%
	In this example the scaled transverse momentum $x=p_T$/(1\,GeV) serves as the input parameter, $0\leq x \leq 2$. For inference, each $p_T$ bin is represented by its midpoint value.}
In the ``truth'' model we use in the particlization step the Grad 14-moment momentum distribution \cite{Denicol:2012cn} for $f_\mathrm{eq}+\delta f$, and we set the switching energy density to $\epsilon_{\rm sw} = 0.2\,$GeV/fm$^3$. For various different examples, we consider $\eta/s$ either as a constant, or as a function of temperature parametrized by \cite{JETSCAPE:2020mzn}
\begin{align}
    \Bigl(\frac{\eta}{s}\Bigr)(T) =&\,(\eta/s)_\mathrm{kink} + \Theta(T - T_{\rm kink}) a_\mathrm{high} (T - T_{\rm kink}) 
    \nonumber \\ \label{etabys_param}
    &+ \Theta(T_{\rm kink} - T) a_\mathrm{low} (T - T_{\rm kink}).
\end{align}

The theoretical model to be compared with these mock data is the same \texttt{iEBE-MUSIC} model but with a different ansatz for the momentum distribution at particlization (assuming local thermal equilibrium $f_\mathrm{eq}$ instead of the Grad 14-moment momentum distributions for the particlized hadrons). For the various examples to be presented, $\eta/s$ is considered either as a constant or parametrized as in Eq.~\eqref{etabys_param}, and at an {\it a priori} unknown switching energy density $\epsilon_{\rm sw}$. Our prior knowledge about the domain of validity of the theoretical model is modeled using the two distinct GP-kernels discussed earlier. Kernel \texttt{I} assumes that the model is more reliable at small $p_T$ than at high $p_T$, while Kernel \texttt{II} assumes that the model is accurate at $p_T=0$ but becomes increasingly less reliable as $p_T$ increases. These choices are motivated by the use of an equilibrium momentum distribution at particlization which omits higher-order $p_T$ corrections associated with shear viscous effects on the particlization hypersurface. In the examples that follow, we perform Bayesian parameter inference to determine posteriors for $\eta/s$ and $\epsilon_{\rm sw}$ under three scenarios: (i) without the MD term (shown in red), (ii) with MD using GP-kernel \texttt{I} (blue), and (iii) with MD using GP-kernel \texttt{II} (green).

%%%%%%%%%%%%%%%%%%%%%%%%%%%%%%%%%%%%%%%%%%%%%%%%%%%%%%%%%%%%%%%%

%--------------------------------
\begin{figure*}[tbh!]
    \centering
    \includegraphics[width=\linewidth]{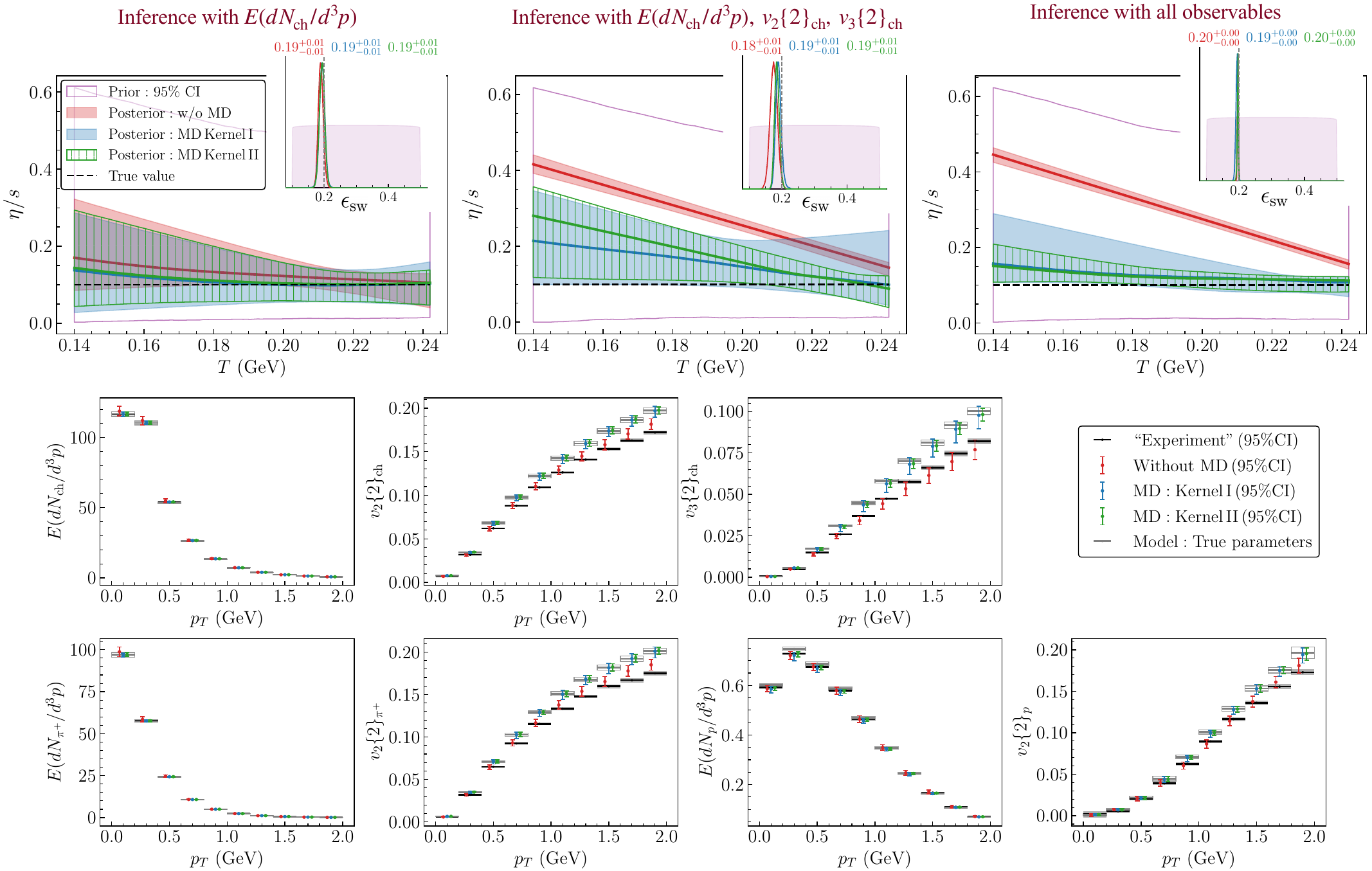}
    \vspace{-6mm}
    \caption{Results for the five-parameter hydrodynamic simulation with parametrized $\eta/s$ using mock data generated with fixed $\eta/s=0.1$ and $\epsilon_{\rm sw}=0.2\,$GeV/fm$^3$. \textbf{Top row:} Plots display the $\eta/s$ posterior (median and $\pm 95\%$ CI) as a function of temperature, with the posterior for $\epsilon_{\rm sw}$ (median $\pm 68\%$ CI) shown in the inset, as additional observables are sequentially incorporated into the Bayesian inference (from left to right). 
    \textbf{Bottom rows:} Model predictions based on inference using all observables.}
    \vspace{-2mm}
    \label{fig:5flat}
\end{figure*}
%--------------------------------

%-------------------------------------------
\subsection{Truth: constant \texorpdfstring{$\eta/s$}{}; two model parameters.}
\label{sec:hydro_case1}
%-------------------------------------------

In this example, mock experimental data are generated using a constant $\eta/s{\,=\,}0.1$ and $\epsilon_{\rm sw}{\,=\,}0.2\,$GeV/fm$^3$. From these data we infer the two model parameters $\eta/s$ and $\epsilon_{\rm sw}$ using a model that ignores viscous corrections during particlization.

As in the ball-drop example, we perform the inference by progressively incorporating additional observables to assess their influence on the inferred model posteriors. The results for this example are shown in Fig.~\ref{fig:2flat}. In the top row, the left panel displays the corner plot when only $E(dN_{\text{ch}}/d^3p)$ is considered, the middle panel adds $v_2\{2\}_{\rm ch}$ and $v_3\{2\}_{\rm ch}$, and the right panel includes all observables.

Using only the $E(dN_{\text{ch}}/d^3p)$ measurements (top left), all three cases (i)-(iii) yield posteriors that are statistically consistent with the truth. However, once the flow observables $v_2\{2\}_{\rm ch}$ and $v_3\{2\}_{\rm ch}$ are included (top middle), in the no-MD case (red) the Bayesian fit clearly struggles to find optimal values for the parameters, producing significantly shifted posteriors that hit the edge of the prior range (``very wrong'') while exhibiting exceedingly small uncertainties (``very confident''). Including MD using Kernel \texttt{I} (blue), the posteriors are once again statistically consistent with the truth, albeit with increased uncertainty (``much less confidence'') for $\eta/s$. With Kernel \texttt{II} (green) the MD term captures the model imperfections for $v_2$ and $v_3$ much better, yielding posteriors similar to those obtained by ignoring the charged hadron anisotropic flow data. Once all available observables are included (top right), the inferred values for both model parameters agree with their true values, within significantly reduced uncertainties, as long as MD is accounted for (blue and green); without MD $\eta/s$ is very confidently inferred to be about $40\%$ larger than its true value.

The bottom panel of Fig.~\ref{fig:2flat} compares the mock experimental data with their predictions from the calibrated model. Black boxes denote the median and $95\%$ credible intervals (CI) of the experimental data, while open gray boxes show predictions from the model using the true parameter values, the difference highlighting the effects of model imperfections. The colored (red, blue, green) points and vertical bars show the median and $95\%$ CI of the model predictions obtained from the inferred posteriors in the top right panel. Overall, predictions with MD (blue and green) are closer to the model predictions with the true parameters while those without MD (red) align more closely with the data. This demonstrates that the MD framework prioritizes obtaining correct parameter estimates over merely fitting the observables.

%%%%%%%%%%%%%%%%%%%%%%%%%%%%%%%%%%%%%%%%%%%%%%%%%%%%%%%%%%%%%%%%

%--------------------------------
\begin{figure*}[tbh!]
    \centering
    \includegraphics[width=\linewidth]{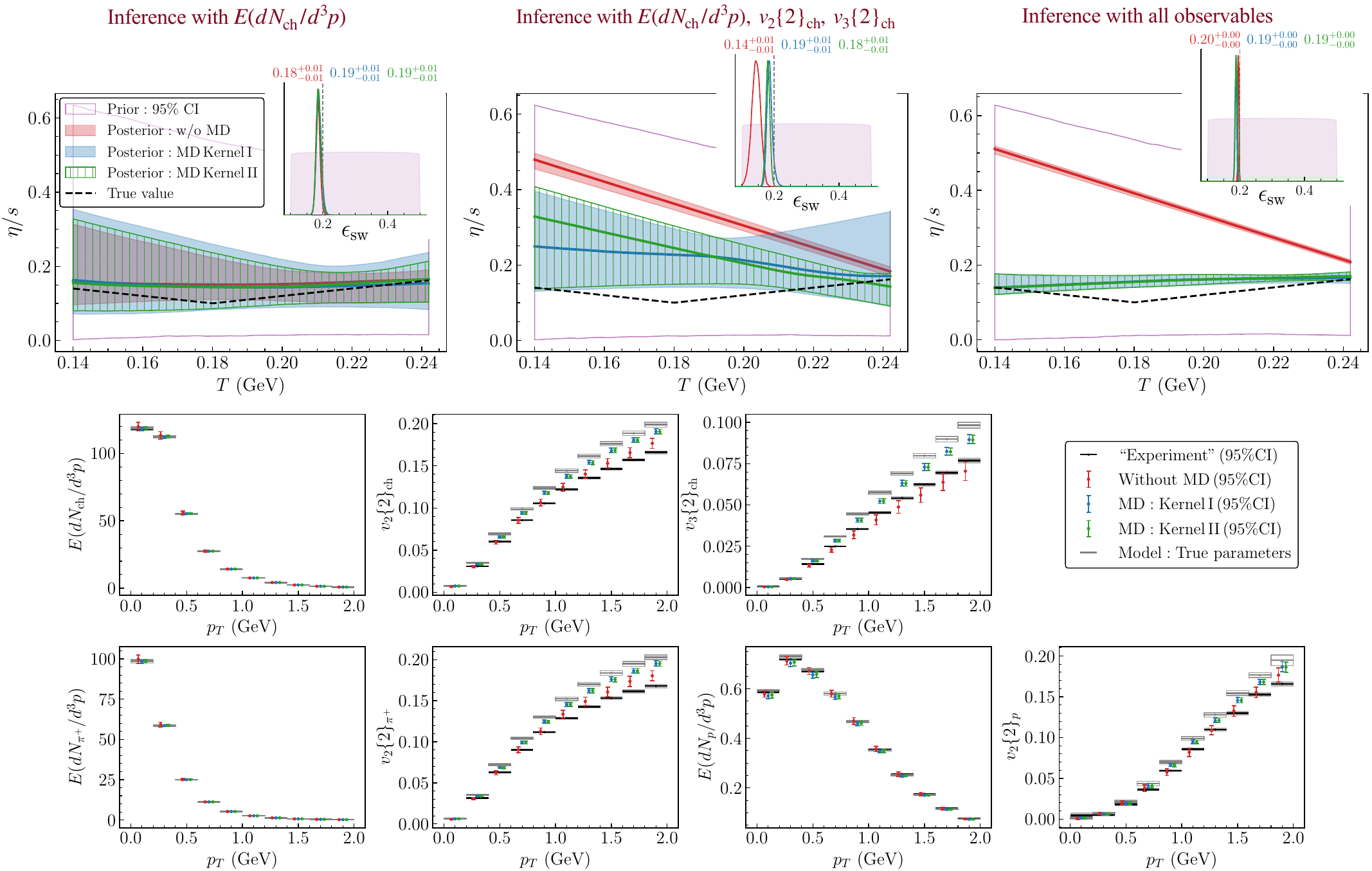}
    \vspace{-6mm}
    \caption{Results for the five-parameter hydrodynamic simulation with parametrized $\eta/s$ using mock data generated using a parametrization of $\eta/s$  with $T_{\rm kink}{\,=\,}0.18\,{\rm GeV}$, $a_{\rm low}{\,=\,}{-}1\,{\rm GeV}^{-1}$, $a_{\rm high}{\,=\,}1\,{\rm GeV}^{-1}$, and $(\eta/s)_{\rm kink}{\,=\,}0.1$, along with $\epsilon_{\rm sw}{\,=\,}0.2\,$GeV/fm$^3$. The plot layouts and legends are identical to those in Fig.~\ref{fig:5flat}. The corner plot for all five model parameters, when all available observables are included, is provided in~\ref{sec:app_corner}, Fig.~\ref{fig:5notflat_modelcorner} .}
    \vspace{-2mm}
    \label{fig:5notflat}
\end{figure*}
%--------------------------------

%-------------------------------------------
\subsection{Truth: constant \texorpdfstring{$\eta/s$}{}; five model parameters.}
\label{sec:hydro_case2}
%-------------------------------------------

Using the same mock experimental data as before we now try to infer from them both $\epsilon_{\rm sw}$ and the temperature dependence of $\eta/s$, using a parametrization of $(\eta/s)(T)$ with four parameters ($T_{\rm kink}$, $a_{\rm low}$, $a_{\rm high}$, and $(\eta/s)_{\rm kink}$)  taken from \cite{JETSCAPE:2020mzn}. The true value of $\eta/s$ is 0.1, without any dependence on $T$. Note that this corresponds to a singular line in the 4-dimensional parameter space at $(\eta/s)_{\rm kink}{\,=\,}0.1$, $a_{\rm low}{\,=\,}a_{\rm high}{\,=\,}0$, and $T_{\rm kink}$ undetermined. As shown in Figure~\ref{fig:5flat}, which we will now discuss, this last point is problematic for Bayesian parameter inference. Similar to Fig.~\ref{fig:2flat}, the top row displays the inferred $\eta/s$ as a function of temperature and (in the inset) the posterior for $\epsilon_{\rm sw}$, with the same color coding as before. The corner plot for all five model parameters, for the case when all available observables are included, is provided in Fig.~\ref{fig:5flat_modelcorner} in~\ref{sec:app_corner}. The bottom row of panels for this 5-parameter model mirrors those in Fig.~\ref{fig:2flat} for the 2-parameter model.

When only the momentum distribution $E(dN_{\text{ch}}/d^3p)$ is considered (top left), all three cases yield posteriors that are statistically compatible with the truth. However, for $\eta/s$ the posteriors accounting for MD (blue and green) overlap significantly better with the true values than those without MD (red). Once $v_2\{2\}_{\rm ch}$ and $v_3\{2\}_{\rm ch}$ are added to the calibration data (top middle), none of the posteriors uniformly covers the truth any more. When not accounting for MD (red), the posterior credible intervals for $\eta/s$ are very narrow (suggesting high confidence) but are completely inconsistent with the truth. Including MD helps a little but still infers $95\%$ credible intervals that do not cover the true value of $\eta/s$ over most of the temperature range. Apparently, the inference algorithm tries to compensate for the model's tendency to over-predict $v_2$ and $v_3$ by increasing $\eta/s$.%
    \footnote{This inability of the MD framework to uniquely identify the contribution of model parameters and GP hyperparameters, respectively, has been discussed in the model discrepancy literature \cite{Kennedy2002, Higdon2004, Bayarri01052007, Arendt:etal2012:1, Arendt:etal2012:2, Brynjarsdottir_2014, GARDNER2021107381}. We note that the framework presented in this work for selecting the kernel differs from earlier approaches, thereby altering the impact of this identifiability issue.}
When all observables (including the spectra and elliptic flows of identified pions and protons) are added to the calibration data (top right), all posteriors become narrower (i.e., more constrained). The overlap of the posteriors accounting for MD improves, whereas the posteriors that do not account for MD degrade further.

The model predictions in the bottom row of Fig.~\ref{fig:5flat} (obtained from the inferred posteriors in the top right panel) reinforce our observations made in the two-parameter case: Predictions that include MD do not attempt to overfit the data (black) but remain closer to the model predictions using the true parameter values (open gray). In contrast, the predictions without MD overfit the data, resulting in the incorrect inference for $\eta/s$ observed in the top right panel.

%%%%%%%%%%%%%%%%%%%%%%%%%%%%%%%%%%%%%%%%%%%%%%%%%%%%%%%%%%%%%%%%

%-------------------------------------------
\subsection{Truth: temperature-dependent \texorpdfstring{$\eta/s$}{}; five model parameters.}
\label{sec:hydro_case3}
%-------------------------------------------

In this example the mock experimental data are generated using a temperature dependent $(\eta/s)(T)$ \cite{JETSCAPE:2020mzn}, with parameters set to $T_{\rm kink}{\,=\,}0.18\,{\rm GeV}$, $a_{\rm low}{\,=\,}{-}1\,{\rm GeV}^{-1}$, $a_{\rm high}{\,=\,}1\,{\rm GeV}^{-1}$, and $(\eta/s)_{\rm kink}{\,=\,}0.1$. The model remains the same as in the preceding case, with a total of five parameters.

The results are shown in Fig.~\ref{fig:5notflat} and support the overall conclusions drawn from the preceding cases: inference including the MD term performs significantly better than inference without MD. Still, we observe that in this last example none of the cases (with or without accounting for MD) accurately captures the shape of $(\eta/s)(T)$. We suspect that the available data do not provide sufficient information to properly constrain it given the theoretical deficiencies (i.e. the neglect of viscous corrections in the particlization routine) of the model. Even with MD, simply knowing that the model is more reliable at small $p_T$ than at large $p_T$ is not enough to resolve the shape of $(\eta/s)(T)$. Nonetheless, the improvement over conventional Bayesian fits remains substantial.

%%%%%%%%%%%%%%%%%%%%%%%%%%%%%%%%%%%%%%%%%%%%%%%%%%%%%%%%%%%%%%%%

%-------------------------------------------
\section{Summary and Outlook}
\label{sec:summary}
%-------------------------------------------

Often the goal of model-data comparison is to extract physically meaningful model parameters, not solely to fit data. In this work, we introduce a framework that explicitly quantifies theoretical uncertainties by statistically modeling theory errors using Gaussian processes, guided by qualitative knowledge of a theory's varying reliability across the input domain. This approach yields more robust and precise parameter estimates, as demonstrated in two different systems. In each case, systematic improvements are observed as additional experimental observables are incorporated into the calibration, leading to results that significantly outperform conventional Bayesian inference without model discrepancy (MD).

The MD framework presented in this Letter opens a new avenue for robust model-data comparison. With readily available qualitative information about a theory's domain, the framework can be used to obtain reliable physical parameter estimates while correctly assessing the model's limitations and avoiding overfitting. Importantly, it allows for the extraction of meaningful information from data that are known {\it a priori} to differ from the theoretical model --- data that are typically discarded in conventional analyses. It also allows models of varying complexity and domains of validity to be directly compared against data. Finally, the framework enables validation of the inferred parameter posteriors in two ways: (i) posteriors should remain consistent as more observables are included in inference, and (ii) results obtained using a kernel with strong prior information (e.g., Kernel \texttt{II}) should be tighter yet consistent with those derived from more conservative modeling assumptions (e.g., Kernel \texttt{I}).

%%%%%%%%%%%%%%%%%%%%%%%%%%%%%%%%%%%%%%%%%%%%%%
\section*{Data availability}
%%%%%%%%%%%%%%%%%%%%%%%%%%%%%%%%%%%%%%%%%%%%%%

The code and data to reproduce these results is publicly available in~\cite{jaiswal_2025_16986775}.

%%%%%%%%%%%%%%%%%%%%%%%%%%%%%%%%%%%%%%%%%%%%%%
\section*{Acknowledgements}
%%%%%%%%%%%%%%%%%%%%%%%%%%%%%%%%%%%%%%%%%%%%%%

We thank Daniel Phillips for useful feedback on the manuscript. S.J. thanks Jean-Paul Blaizot, Chandrodoy Chattopadhyay, Yi Chen, Sangyong Jeon, and Jean-Yves Ollitrault for insightful discussions and feedback. This research was supported by the CSSI program Award OAC-2004601 (BAND collaboration \cite{BAND_Framework}) (S.J., R.J.F., U.H, M.T.P.). C.S. is supported by the U.S. Department of Energy (DOE), Office of Science, Office of Nuclear Physics, under DOE Award No. DE-SC0021969 and DE-SC0024232. C.S. acknowledges a DOE Office of Science Early Career Award. R.J.F. was supported in part by the National Science Foundation under award PHY-2209442. R.J.F. also acknowledges support from the ExtreMe Matter Institute EMMI at the GSI Helmholtzzentrum f\"ur Schwerionenforschung GmbH, Darmstadt, Germany. Heavy-ion simulations were done using services provided by the OSG Consortium \cite{osg07,osg09,osg77,osg78}, which is supported by the National Science Foundation awards $\#2030508$ and $\#2323298$. Bayesian inference was done on resources provided by the Ohio Supercomputer Center \cite{OhioSupercomputerCenter1987} (Project No. PAS0254).

% %%%%%%%%%%%%%%%%%% Bibliography %%%%%%%%%%%%%%%%%%
% \bibliographystyle{elsarticle-num}
% \bibliography{ref}
% %%%%%%%%%%%%%%%%%%%%%%%%%%%%%%%%%%%%%%%%%%%%%%%%%%

\clearpage
\onecolumn
\appendix

%-------------------------------------------
\section{Corner plots of model parameters for heavy-ion simulation}
\label{sec:app_corner}
%-------------------------------------------

\begin{figure}[tbh!]
    \centering
    \includegraphics[width=\linewidth]{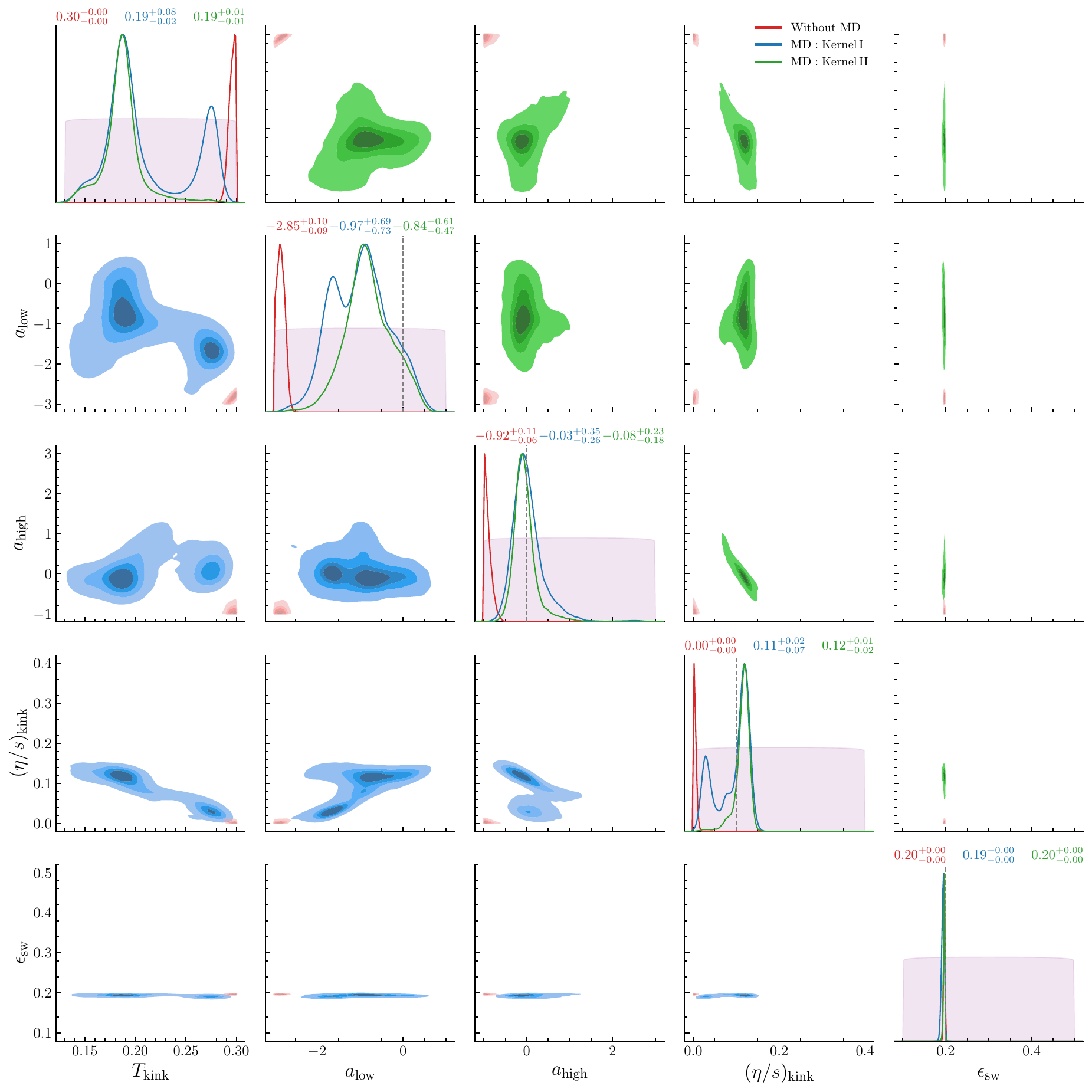}
    \vspace{-6mm}
    \caption{Corner plot of all model parameters (with median $\pm 68\%$ CI displayed on the diagonals) for the five-parameter model with parametrized $\eta/s$, using mock data generated with fixed $\eta/s=0.1$ and $\epsilon_{\rm sw}=0.2\,$GeV/fm$^3$, corresponding to the results shown in Fig.~\ref{fig:5flat} in the main text.}
    \vspace{-2mm}
    \label{fig:5flat_modelcorner}
\end{figure}

We show the corner plot of all model parameters corresponding to Fig.~\ref{fig:5flat} in the main text in Fig.~\ref{fig:5flat_modelcorner}. Overall, the posteriors agree better with the true values (indicated by dashed gray vertical lines) when the model discrepancy (MD) term is included (for both Kernel \texttt{I} and Kernel \texttt{II}) compared to the no-MD case (red). Note that, since the generated mock experimental data do not depend on the parameter $T_{\rm kink}$, its posterior should ideally mirror the prior (shaded purple region). However, we observe that the posterior for $T_{\rm kink}$ peaks around a certain value even in the MD inference, although its width is significantly larger than in the no-MD case. This may be an artifact of the overly flexible model, where the MD cases struggle to disentangle the effects of the model parameter from the GP hyperparameters. Nonetheless, the resulting $\eta/s$ posterior shown in Fig.~\ref{fig:5flat} in the main text does not appear to depend crucially on this parameter.

%--------------------------------
%--------------------------------

\begin{figure}[tbh!]
    \centering
    \includegraphics[width=\linewidth]{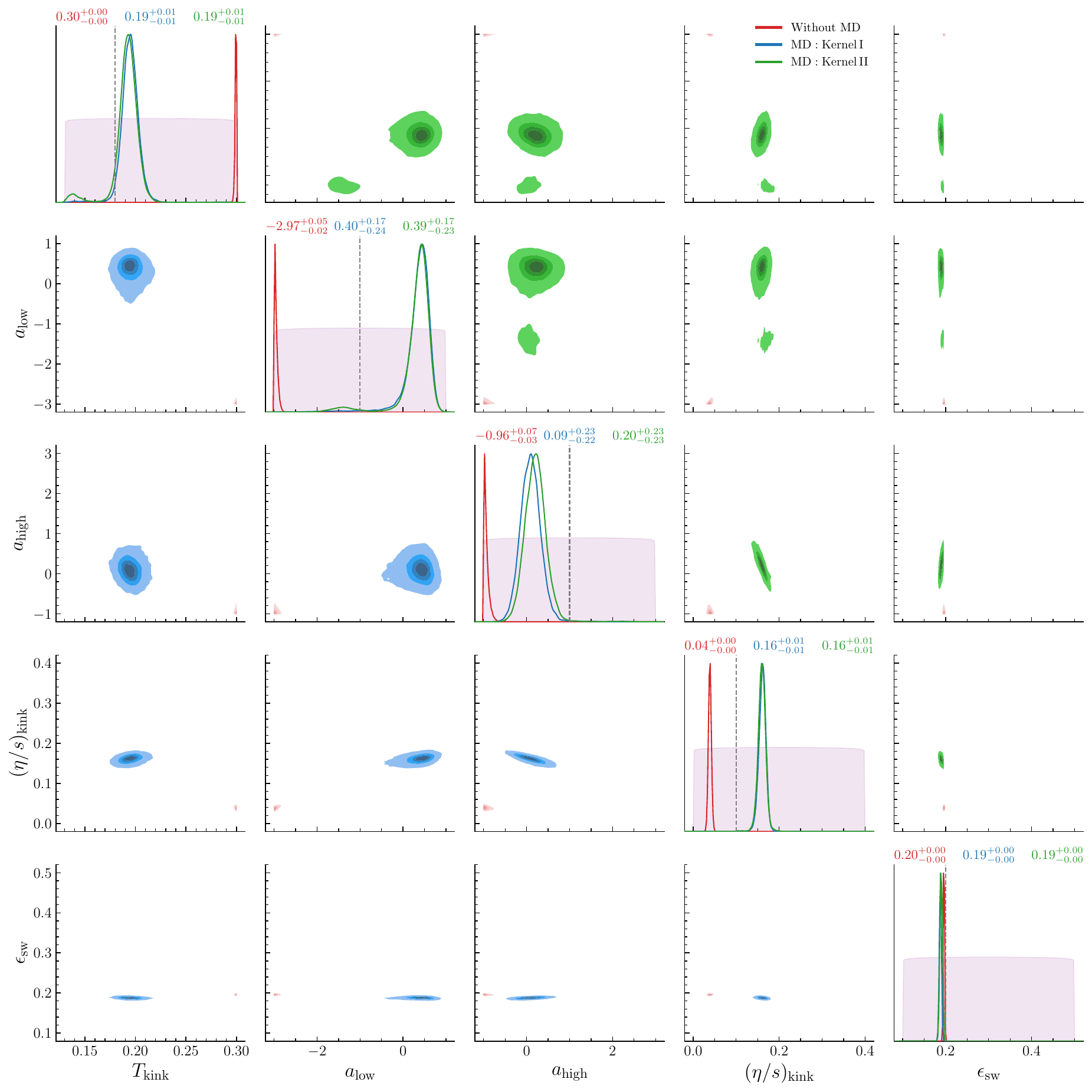}
    \vspace{-6mm}
    \caption{Corner plot of all model parameters (with median $\pm 68\%$ CI displayed on the diagonals) for the five-parameter model with parametrized $\eta/s$, using mock data generated via the parametrization with $T_{\rm kink}=0.18\,{\rm GeV}$, $a_{\rm low}=-1$, $a_{\rm high}=1$, and $(\eta/s)_{\rm kink}=0.1$, along with $\epsilon_{\rm sw}=0.2\,$GeV/fm$^3$. This plot corresponds to the results shown in Fig.~\ref{fig:5notflat} in the main text.}
    \vspace{-2mm}
    \label{fig:5notflat_modelcorner}
\end{figure}

The corner plot of all model parameters corresponding to Fig.~\ref{fig:5notflat} in the main text is shown in Fig.~\ref{fig:5notflat_modelcorner}. Overall better agreement with the true values is observed when the model discrepancy (MD) term is included (blue, green) compared to the no-MD case (red). The inability of the posteriors to accurately capture the slopes $a_{\rm low}$ and $a_{\rm high}$ is reflected in the nearly flat shape of $\eta/s$ as a function of temperature in Fig.~\ref{fig:5notflat} for the MD cases. We observe that the absence of increasing specific viscosity on either side of $T_{\rm kink}$, as seen in the truth, is compensated by an overestimated posterior for $(\eta/s)_{\rm kink}$ (for blue, green). We attribute the framework's inability to accurately capture the posteriors for these parameters to a combination of limitations in the theoretical model, insufficient information in data to constrain the slopes, and weak prior information regarding the model's domain used in the kernel modeling.

%--------------------------------
%--------------------------------

\twocolumn

%%%%%%%%%%%%%%%%%% Bibliography %%%%%%%%%%%%%%%%%%
\bibliographystyle{elsarticle-num}
\bibliography{ref}
%%%%%%%%%%%%%%%%%%%%%%%%%%%%%%%%%%%%%%%%%%%%%%%%%%

\end{document}